\documentclass[prl, twocolumn,amsmath,amssymb]{revtex4}
\usepackage{graphicx}
\usepackage{color}

\begin{document}

\title{The very long range nature of capillary interactions in liquid films}

\author{R. Di Leonardo$^1$, F. Saglimbeni$^2$ \& G. Ruocco$^2$}

\affiliation{
$^1$ INFM-CRS SOFT c/o Universit\'a di Roma ``La Sapienza'', I-00185, Roma, Italy\\
$^{2}$Dipartimento di Fisica, Universit\'a di Roma ``La Sapienza'', I-00185, Roma, Italy}

\date{\today}

\begin{abstract} 
Micron-sized objects confined in thin liquid films interact through forces
mediated by the deformed liquid-air interface. This capillary interactions
provide a powerful driving mechanism for the self-assembly of ordered
structures such as photonic materials or protein crystals. Direct probing of
capillary interactions requires a controlled force field to independently
manipulate small objects while avoiding any physical contact with the
interface.  We demonstrate how optical micro-manipulation allows the direct
measurement of capillary interactions between two micron sized spheres in a
free standing liquid film.  The force falls off as an inverse power law in
particles separation.  We derive and validate an explicit expression for this
exponent whose magnitude is mainly governed by particles size. For micron-sized
objects  we found an exponent close to, but smaller than one, making capillary
interactions a unique example of strong and very long ranged forces in the
mesoscopic world. 
\end{abstract}

\maketitle It is a well known fact that small objects floating on a liquid
surface cluster together. Bubbles on the surface of a soap solution\cite{bragg}
or cereals in a bowl of milk\cite{cheerio} attract each other with long range
forces arising from the interface deformation under particles weight (or
buoyancy for bubbles). Close packed configurations for such macroscopic objects
are found to minimize gravitational potential energy. Shrinking lengths to the
mesoscopic scale, particles weight soon become too weak to produce any
significant deformation and hence attraction.  However if the  particles are
confined in thin liquid films a deformation of the interface is unavoidable.
This is the case, for example, of a colloidal suspension drying on a solid
substrate, or dispersed in a free standing thin film. When the thickness of the
liquid film becomes smaller than the bead diameter, the interface has to deform
with an increase in surface energy. The liquid interface will then react on the
particles with forces aiming to reach a minimum surface (energy) configuration,
that usually corresponds to close packed two dimensional crystals\cite{2d}.
Such phenomena, already observed by Perrin in 1909\cite{perrin}, have attracted
considerable interest in recent times due to their relevance for the
engineering of photonic materials\cite{vlasov} and protein
crystallography\cite{yoshimura}.  Consequently, a strong effort has been
devoted to the theoretical analysis of the involved forces, resulting in a long
series of papers reviewed in\cite{review}. Prediction for macroscopic objects
have been confirmed by experiments on immersed cylinders \cite{velev} or
particles attached to holders\cite{dushkin}.

However no experiment so far has been able to directly measure the strong
capillary force acting between an isolated pair of mesoscopic objects, despite
the fact that it is in the mesoscopic and nanoscopic realm that this effect
finds the most interesting applications.  Any physical contact with the
particles would inevitably produce a significant deformation of the liquid-air
interface and dramatically affect the interaction. On the other hand, due to
long ranged hydrodynamic interactions in 2D, particles mobility are very
sensitive to interparticle distances and force measurements are difficult to
deduce from particles trajectories. A static, highly non-invasive method is
required for a direct and reliable measurement of these interactions.

In this Letter we demonstrate how optical micro-manipulation \cite{ashkin}
allows the precise measurement of capillary interactions between two micron
sized spheres confined in a free standing thin liquid film.  Holographic
optical tweezers \cite{hot} allow to isolate a single pair of particles and
scan their relative distance from close contact to several tens of diameters.
Capillary forces will tend to push the particles towards each other and out of
the optical traps until the restoring trap forces will balance the attractive
force.  The intensity of capillary force can be then deduced by measuring
particles displacements from trap centers, after trap calibration.  Working in
a free standing liquid film is essential for accurate capillary force
determinations since no particle-substrate interactions have to be taken into
account.

\begin{figure}
\includegraphics[width=.45\textwidth]{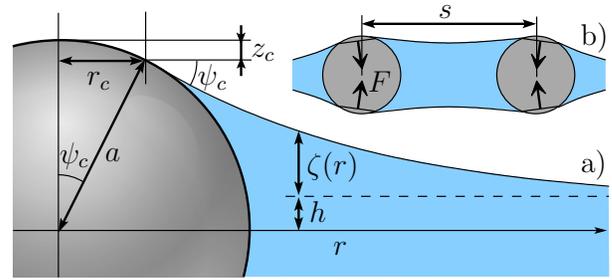}
\label{fig1}
\caption{
(Color online) Wetting geometry. Complete wetting is assumed, so that the liquid film wraps
the particle inside a spherical cap of radius $r_c$. For $r>r_c$ the liquid-air
interface is free-standing and slowly falls to the large distance height $h$.
The local height of the interface measured from $h$ is indicated by the
function $\zeta(r)$, whose gradients are assumed to be small everywhere
($\psi_c\ll1$).}
\end{figure}

The force law can be anticipated by calculating the surface tension forces
acting on a pair of spherical objects confined in a thin liquid film.
Kralchewsky et al.\cite{kral} derived the shape of the meniscus around the two
particles using the method of matched asymptotic expansions. An implicit
expression for the force law was obtained, whose evaluation required the
numerical solution of a system of nonlinear equations in the accessible system
parameters. Confining here to mesoscopic objects, we show that Kralchewsky
solution for the force can be very well approximated by an inverse power law
whose parameters have an explicit expression in terms of the system's physical
properties.  We assume complete wetting (zero contact angle) with the liquid
surface departing from a contact ring of radius $r_c$ with continuous slope
angle $\psi_c$ (Fig.  1a).  Surface tension will exert a force on the contact
ring whose resultant is orthogonal to the ring plane and has a modulus:

\begin{equation} \label{F} F=\gamma\; 2\pi r_c\ \sin\psi_c \end{equation}

where $\gamma$ is the liquid-air surface tension ($\gamma\simeq35$ mN/m in our
experiment).  For small gradients $r_c=a \sin\psi_c\sim a \psi_c$,
$z_c=a(1-\cos\psi_c)\simeq a\psi_c^2/2$, and the force $F\simeq 4\pi\gamma z_c$
is then proportional to $z_c$ with a strength of order 1 nN/nm.  The order of
magnitude of other forces into play is 1 pN for particle's weight and 100 pN
for the maximum optical force exerted by our trap. 
The interface is supposed to be flat and horizontal far from the particle and
we indicate with $\zeta(r)$ the local vertical displacement of the interface
from this reference surface.  In the small gradient approximation, Laplace
equation for the pressure drop across free standing portion of the upper
surface ($r>r_c$) reads\cite{landau}:

\begin{equation}
\label{laplace}
\mathbf \gamma\nabla^2\zeta(r)=\rho g \zeta(r)
\end{equation}

where $\rho$ is the liquid density and $g$ the acceleration due to gravity.
The only axisymmetrical solution to (\ref{laplace}) vanishing at infinity takes
the form\cite{nico}:

\begin{equation}
\zeta(r)=-\frac{\tan\psi_c}{q}\frac{K_0(q r)}{K_1(q r_c)}
\end{equation}

where $K_i(x)$ is the modified Bessel function of ith order\cite{abramowitz}
and $q^{-1}=\sqrt{\gamma/\rho g}$ is the capillary length. This is the length
scale below which gravitational forces play no role in determining the
interface shape and it's about $2$ mm in  typical solvents.  Therefore we can
safely replace $K_0(x)$ and $K_1(x)$  with their small argument expansion and
write for small gradients:

\begin{equation} 
\label{profiloth}
\zeta(r)=-2 z_c \left[\log(q r/2)+\gamma_e \right]
\end{equation}

where $\gamma_e$ is the Euler-Mascheroni constant.  At the micron scale
gravitational energy will be negligible and the bottom surface will have a
symmetrical shape to the top one. Therefore two equally strong capillary forces
will act on the top and bottom contact rings of an isolated particles but they
will cancel out giving no net force.  When a second particle is introduced in
the film at a distance $s$, it will in turn contibute to the interfaces
deformation producing a tilt of the contact lines around the first particle
(Fig.  1b). For small deformations we can still think of contact lines as
circles but this time slightly tilted.  Within the superposition approximation
by Nicolson\cite{nico}, the amount of tilt would simply be given by the
gradient of the surface deformation produced by an isolated particle at the
location of the second one. According to (\ref{profiloth}) the interface height
around an isolated particle is expected to decay logarithmically with the
distance producing a tilt of the contact lines that falls off as the inverse
interparticle distance.  Capillary forces $F$ acting on the contact rings will
not be balanced anymore but a net attractive force would appear whose intensity
is given by the projections of $F$ on the film reference plane:

\begin{equation}
\label{force}
f(s)=2 F(s) \left.\frac{\partial \zeta(r)}{\partial r}\right|_{r=s}=
16\pi\gamma z_c^2(s) \frac{1}{s}
\end{equation}

where $s$ is the interparticle distance.  The resulting force is than
proportional to the square of $z_c$ and inversely proportional to the distance
separating the particles.  In analogy with Coulomb electrostatic force in 2D,
the quantity $z_c$ is usually called the ``capillary charge'' of the
particle\cite{charge}.  However, the capillary charge is not an intrinsic
constant property but a slowly varying function of particle separation. This
function can be calculated by imposing the continuity and differentiability of
the interface across the contact ring.

\begin{eqnarray}
\label{boundary1}
&&\zeta(r_c)+\zeta(s)+h=a-z_c\\
\label{boundary2}
&&r_c=a \sin\psi_c\sim\sqrt{2 a z_c} 
\end{eqnarray}

where we rely on the  superposition approximation \cite{nico} to express the
interface vertical displacement field as the sum of two single particle
displacement fields (\ref{profiloth}).  We are also assuming in
(\ref{boundary1}) that the deformation field $\zeta(s)$ produced by one
particle is constant over the contact line of the other.  Equations
\ref{boundary1},\ref{boundary2} can be solved analytically giving:

\begin{equation}
\label{zc}
z_c(s)=\frac{a-h}{-W\left[-q^4 s^2 a(a-h) \exp[4\gamma_e-1]/8\right]}
\end{equation}

Where $W$ is the Lambert-W function \cite{lambert}. The above expression for
$z_c$ is a slowly varying function of $s$ in mesoscopic systems ($a<s\ll
q^{-1}$) and it can be well approximated by its logarithmic expansion about
$s=2a$:

\begin{eqnarray}
&&z_c(s)\sim z_{0} (s/2a)^{\alpha}\\[.2cm]
&&z_0=z_c(2 a)\\[.1cm]
&&\alpha=\left.\frac{d\log z_c}{d\log s}\right|_{s=2 a}=\frac{2 z_{0}}{a-h-z_{0}}
\end{eqnarray}

Accounting for changes in $z_c$ the attractive force (\ref{force})
will still display a power law behavior but with an exponent smaller than one:  

\begin{equation}
\label{force0}
f(s)=\frac{16\pi\gamma z_{0}^2}{2a} \left(\frac{2a}{s}\right)^{1-2\alpha}
\end{equation}

where $\alpha$ is an explicit function of the three lengths: particle size,
film thickness, capillary length of the solvent.  The exponent $1-2\alpha$ is
equal to $1$ for $h\rightarrow a$ but then quickly drops to a fairly constant
value as soon as $h$ is small enough to produce significative forces ($\sim$ 1
pN $\sim 10^3 K_B T/a$). This constant value depends practically only on the
particle size and varies very little when using different typical solvent
properties.  Changing particle size from 10 nm to 10 $\mu$m produces a
corresponding exponent variation in the range 0.92 to 0.82.  For our particle
size we predict an exponent of 0.86.

\begin{figure}[ht]
\includegraphics[width=.35\textwidth]{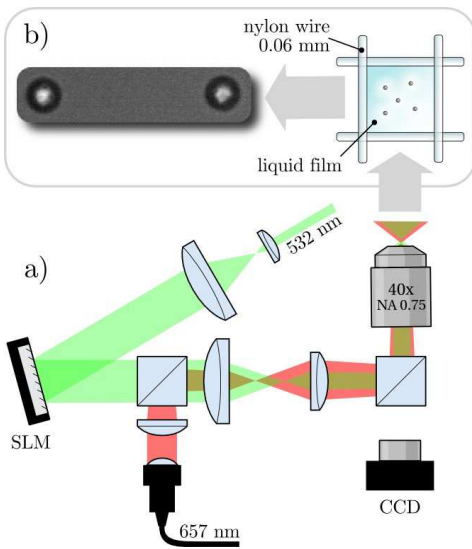} 
\label{fig2}
\caption{ 
(Color online) Schematic view of the experimental setup described in text.
}
\end{figure}

\begin{figure}
\includegraphics[width=.45\textwidth]{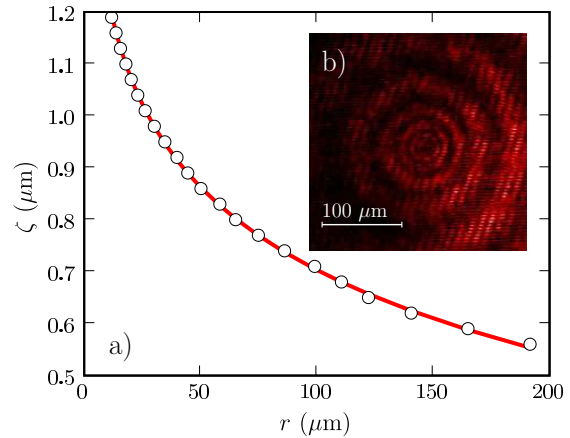}
\label{fig3}
\caption{ 
(Color online) Film thickness profile around an isolated particle. b) Monochromatic light
reflected off the film displays interference fringes centered around an
isolated trapped particle. A period in the fringe pattern corresponds to 240 nm
thickness variation. Figure a) reports as open circles the extracted film
height $\zeta(r)$ versus distance from particle center $r$ together with a
logarithmic fit (red line).}
\end{figure}

A schematic view of the experimental setup is reported in Fig. 2a.  

Latex beads (5.7 $\mu m$ diameter) are dispersed in a 2/3 water - 1/3 glycerol
mixture with added 0.2\% wt surfactant (SDS). A thin film is obtained by
sweeping the solution on a square frame (10 mm side) of nylon wires (60 $\mu m$
thickness). The free standing liquid film is enclosed in a humidity chamber and
placed over the 40x objective of an inverted optical microscope (Nikon
TE2000-U). The same objective is used to focus the laser beam ($\lambda$=532
nm) reflected off a spatial light modulator (Holoeye LCR-2500) into two,
dynamically reconfigurable, optical traps \cite{hot}.  Axial confinement is
guaranteed by the normal components of capillary forces.  We can also access
the thickness variations around an isolated particle by viewing the film under
reflected monochromatic light.  To this end a red diode laser beam, overlapped
to the trapping green beam, is focused by the same microscope objective far
from the film surface.  The observed portion of the red beam wavefront is
approximately plane and the film reflectance than varies with $\cos[4 \pi
\zeta(r)n/\lambda]$ giving interference rings. $n=1.37$ is the refractive index
of the liquid mixture and $\lambda=657$ nm the red laser wavelength.
The distorted film will then show ring shaped
interference fringes centered on the trapped particle (Fig. 3b).  The thickness
profile around an isolated trapped particle can be extracted (within an
additive constant) and is reported here in Fig. 3a as open circles. A clean
logarithmic shape is found up to 200 $\mu$m, in perfect agreement to prediction
in (\ref{profiloth}),  as shown by the logarithmic fit curve (red
line).

To extract the force law one of the two traps is held fixed while the other is
continously scanned through different distances with a step of 2 $\mu$m.  An
image of the two particle is digitally recorded (Fig. 2b) for every scan
step and subsequently used to extract particle positions.  Due to liquid
drainage, the film is slowly but constantly thinning.  When the thickness is so
small that optical forces cannot balance the capillary interaction, one of the
two particles jumps out of the trap and collapses onto the other. Until that
time we can extract the attractive capillary force from the interparticle
distance.  Calling $k_1$ and $k_2$ the two trap elastic strengths, each
particle will be displaced towards the other by a distance $\Delta x_i=f/k_i$.
The observed interparticle separation will be then smaller than trap separation
by an amount:
$\Delta s=(\Delta x_1+\Delta x_2)=f(1/k_1+1/k_2)=f/k'$.
Particle distance will actually fluctuate due to Brownian motion with a mean
squared value given by \cite{meiners}:
$\langle\Delta d^2\rangle=K_B T/k'$
which can be used to experimentally determine $k'=29$ pN/$\mu$m.

\begin{figure}[ht] \includegraphics[width=.45\textwidth]{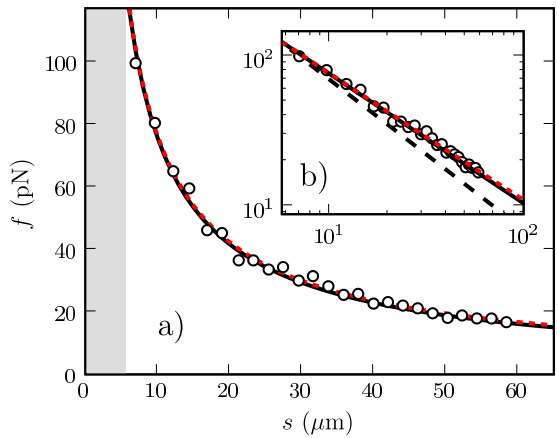} \label{fig4}
\caption{
(Color online) Intensity of the capillary interaction between two beads confined in a thin
film, as a function of their relative distance. Open circles are experimental
determinations.  The black solid line represents a fit to the predicted power
law (\ref{force0}) having the film thickness $h$ as the only free parameter.
Red dashed line is the full Kralchewsky theory\cite{kral} for the same $h$
parameter.  The gray region marks the unaccessible range of separations $s$
smaller than close contact distance $2a$.  Double log plot insert more
immediately shows that the data are well represented by a power law.  A 1/r law
is also reported there, for reference, as a black dashed line.} \end{figure}

As already discussed, the logarithmic film shape in Fig. 3 would lead,
in the Nicolson approximation, to an attractive force decaying roughly as the
inverse of interparticle distance. More precisely we would expect to find the
power law in (\ref{force0}).  Indeed a clean power law is found
experimentally as evidenced by the double log plot reported in Fig. 4b.  The
overall behavior is very well fitted by the power law in (\ref{force0})
leaving the film thickness at large distance $h$ as the only free parameter. We
find $h=2.2\mu\textrm{m}=0.8 a$ that makes an average gradient $\psi_c=0.12$
confirming that the small gradient approximation is justified for our system.
For this fitted  $h$ value we get a power law exponent $1-2\alpha=0.86$. The
full Kralchewsky prediction for the same parameter values, also reported in
Fig. 4b, is almost completely overlapped to our power law expansion, confirming
the goodness of our approximations.

We have shown how optical micromanipulation provides an unprecedented tool for
investigating capillary forces which govern aggregation and self-assembly of
colloids in liquid films.  We provide a static force measurement of the
capillary attraction between an isolated colloidal pair and perform a direct
test of the theoretically predicted power law. The exponent of the power law is
found to be close to, but smaller than one, making capillary forces a quite
unique example of very long ranged interactions in the mesoscopic world.  The
experiment opens the way to a variety of further developments addressing the
role of many body-effects, membrane elasticity, wetting properties, surfactant
dynamics, hydrodynamic interactions in 2D. A deep insight into the nature of
interface-mediated forces at the mesoscopic scale could suggest new routes to
self-assembly of meso- and nano-structures\cite{whitesides}. Optical trapping
of colloidal particles bound to lipid membranes \cite{koltover}  could also
provide new insights in the dynamics of biomembrane inclusions\cite{reynwar}.

\end{document}